\documentclass[aps,prb,superscriptaddress,showpacs,twocolumn]{revtex4}

\usepackage{epsfig}
\usepackage{subfigure}
\usepackage{times}
\usepackage{amsmath}
\usepackage{amsthm}
\usepackage{bm,amssymb,verbatim}
\usepackage{amsbsy}

\begin{document}
\title{Controlling non-Abelian statistics of Majorana fermions in semiconductor nanowires}

\author{Jay D.~Sau}
\affiliation{Condensed Matter Theory Center and Joint Quantum Institute, Department of Physics,
University of Maryland, College Park, Maryland 20742-4111, USA}
\author{David J.~Clarke}
\affiliation{Department of Physics and Astronomy,
University of California, Riverside, CA 92521, USA}
\author{Sumanta Tewari}
\affiliation{Department of Physics and Astronomy, Clemson University, Clemson, SC 29634}

\begin{abstract}
Under appropriate external conditions a semiconductor
 nanowire in proximity to an $s$-wave superconductor can be
in a topological superconducting (TS) phase. This phase supports localized
 zero-energy Majorana fermions at the ends of the wire. However, the non-Abelian exchange
statistics of Majorana fermions is difficult verify because of the one-dimensional
topology of such wires. In this paper we propose a scheme to transport Majorana fermions between the
ends of different wires using tunneling, which is shown to be controllable by gate voltages.
 Such tunneling-generated hops of Majorana fermions can be used to exchange the Majorana fermions.
The exchange process thus obtained is described by a non-Abelian braid
 operator  that is uniquely determined by the well-controlled microscopic
 tunneling parameters.
\end{abstract}

\maketitle

\section{Introduction}
\label{Sec:Intro}

Majorana fermions (MFs) have been the subject of intense recent study in
 part due to their potential application in topological quantum computation
 (TQC).\cite{Kitaev,nayak_RevModPhys'08,Freedman,freedman_larsen_wang,
Wilczek-3,bravyi_kitaev1,bravyi} Unlike ordinary fermionic or bosonic
 operators for which the particle creation operators are the hermitian conjugate
of the annihilation operators, MF operators, $\gamma$, are self-adjoint
 ($\gamma^{\dagger}=\gamma$). In this sense, MFs are
 their own anti-particle and the realization of such excitations 
would be the first example of such particles which had been proposed more than seventy-five years 
ago.\cite{majorana} MFs are of interest for TQC because
 despite having no internal degrees of freedom individually, a pair of
 MFs, say $\gamma_1$ and $\gamma_2$, have two distinct possible states (fusion channels).
 These states, which may be thought of as the two possible occupation states of the complex fermionic 
operator $c^\dagger=\frac{\gamma_1+i\gamma_2}{2}$, are energetically degenerate to a degree exponential in the separation
 of the Majorana fermions, and correspond to the eigenstates of the combined
 operator $\imath\gamma_1\gamma_2$, with eigenvalues $\pm 1$. Such a 
topological protected degeneracy has yet to be seen in nature, and the observation of such would 
be a major break-through in physics. The central idea of TQC is to use the 
topologically degenerate states of a pair of MFs as a 2-level topological qubit which would in principle be protected 
from decoherence.  The  manipulation of the 
 information contained in the topological qubits requires the use of topological braid operations 
 which consists of  moving the MFs around one another.

In the past few years, topological superconductors have
become promising candidates for realizing MFs. \cite{volovik,kopnin,read_green,Zhang,fu_prl'08,schnyder,Sato-Fujimoto,Kitaev1}
Recently, it has been proposed that a semiconductor thin film with Rashba-type spin-orbit
 (SO) coupling together with proximity-induced superconductivity and
Zeeman splitting would be a suitable platform for realizing a Majorana-fermion-carrying topological superconducting
 (TS) state. \cite{sau1,Ann,long-PRB,alicea}
The $s$-wave superconducting pairing potential can be induced in the
semiconductor system by placing it in proximity to a conventional
superconductor such as aluminum. The Zeeman splitting can similarly be induced, in principle, by proximity to a 
magnetic-insulator. \cite{long-PRB}
 The one-dimensional version of this system, i.e. a semiconducting
 nanowire with proximity-induced $s$-wave superconductivity,
 has also been shown to host MF as zero-energy modes
at the ends of the wire under appropriate conditions. \cite{unpublished,roman}
 The one-dimensional nanowire geometry has the specific advantage that the Zeeman splitting
 $V_Z$ in the nanowire is not required to be proximity-induced from a magnetic insulator,
but instead can be introduced by a magnetic field parallel to the nanowire. \cite{roman}
Such a parallel magnetic field would not introduce unwanted orbital effects such as 
vortices if a thin-film superconductor is used to generate the superconducting proximity-effect.
 The proposed semiconducting structures  exist in a TS phase and supports MFs at its ends when the $s$-wave superconducting
 pair potential $\Delta$, Zeeman splitting $V_Z$, and the chemical potential $\mu$  satisfy the condition
 $V_Z^2 > \Delta^2 + \mu^2$. \cite{sau1,long-PRB,roman} Thus the chemical potential $\mu$,
which can be controlled by an external gate potential, can be used to
tune a nanowire from the TS phase to a non-topological (NTS) phase.
  In fact, the $s$-wave proximity effect on a InAs quantum wire, which also has a
 sizable SO coupling, may have already been demonstrated in experiments. \cite{doh}
These semiconductor based proposals for realizing a TQC platform
can take advantage of the considerably advanced semiconductor
fabrication technology.  Therefore, it seems that a Majorana-carrying
 TS state in a semiconductor quantum wire may be within experimental reach.

 Until recently, motivated by experiments, most discussions of observing non-Abelian statistics
 using MFs has been restricted to 2D systems. In 2D non-Abelian systems, the quantum
 information associated with MFs can be manipulated in a topologically
 protected manner by exchanging the Majorana bound states (e.g., by adiabatically moving vortices in \emph{p+ip}
 superconductors).\cite{Ivanov,stone} The protection of the topological degeneracy associated with 
MFs requires the MFs to remain spatially separated at all steps of the exchange. Therefore, at first glance, 
it appears that it is impossible to exchange the MFs at the ends of a 1D wire, since any such attempt would 
necessarily lead to the spatial overlap of MFs at some stage of the overlap process.
 A solution to this problem has been provided by  by Alicea {\it et al.} \cite{alicea1}, who have shown that connecting 
up a system of nanowires into a network allows one to exchange MFs. In their proposal, the two MFs, $\gamma_1$ and $\gamma_2$,
 at the ends of a wire can be exchanged by introducing 
 an additional nanowire $B$. The additional nanowire, $B$, allows one to temporarily move one of the MFs, say $\gamma_1$,
away from the original nanowire, $A$, so that the other MF $\gamma_2$ can be moved across the 
 wire $A$ without colliding with $\gamma_2$. The  MF $\gamma_1$ can then be returned from the wire $B$ back
 to it's original wire $A$.
While this  scheme solves the basic problem of non-Abelian statistics in $1$D, 
 it requires the transport of a MF across a tri-junction between
 two topological nanowires $A$ and $B$, which is potentially a more 
complex topological object than the simple topological nanowire. 
The continuous transport of MFs through such a tri-junction is potentially
 dependent on details of the junction that may be difficult to control.\cite{david}
 Moreover, from a theoretical point of view, the explicit determination of the non-Abelian 
statistics in this geometry in terms of the microscopics of the junction is somewhat complicated.

In this paper we propose an alternative scheme to transport MFs at the ends of one dimensional
 semiconductor nanowires
 where the ends of the nanowires remain fixed, but the tunneling amplitudes between the end MFs
 are varied. Bringing the end MFs closer together allows one to create a non-vanishing Hamiltonian
of the MFs which can generate effecting MF hopping from one end site to another.
 Using this picture of a dynamically changing tunneling Hamiltonian,
 we will be able to derive a simple explicit expression for the non-Abelian statistics transformation of the MFs 
in terms of tunneling matrix elements.

\section{Outline and summary of results}
 As mentioned in the previous paragraph, MFs are strictly zero-energy modes 
with an associated topological degeneracy only in the limit when they are separated by a distance that is large 
compared to the decay length of the MFs. The two states of the MFs $\gamma_1$ and $\gamma_2$ can be described in terms of the 
2 possible occupation states of the Dirac fermion $c^\dagger=\gamma_1+i\gamma_2$. These 2 states correspond to the eigenvalues 
$0$ and $1$ of the number operator $n=c^\dagger c=\frac{1+i\gamma_1\gamma_2}{2}$. In general, the Hamiltonian for a pair of MFs
with a non-negligible splitting produces a splitting between the 2 energy states and can be written as 
\begin{equation}
H_{tunneling}=i\zeta_{12}(x)\gamma_1\gamma_2\label{tunneling}
\end{equation}
where $\zeta_{12}(x)$ is the tunneling matrix element for the MFs which depends
 on the separation $x$ between the MFs $\gamma_1$ and $\gamma_2$.\cite{stone}
 The energy splitting between the $n=0$ and $n=1$ states is given by $|\zeta_{1,2}(x)|$.
Therefore, the topological degeneracy of MFs  emerges only in the limit $x\gg \xi$ when the 
MF overlap matrix-element $\zeta_{12}(x)$ vanishes because of the
 localization of the MF wave-functions.
Here $\xi$ is the localization length of the MFs.
 The tunneling of MFs at the ends of different wires, whose ends
 are placed close together, is entirely analogous to the
 tunneling of electrons between two quantum dots which can be controlled by raising and lowering the
 barrier between the dots. Similarly, tunneling amplitudes between MFs on different semiconductor nanowires can
 be controlled simply by adding a gate controllable tunnel barrier between the MFs. \cite{universal}
  Gate voltages can also induce tunneling between MFs at the ends 
of the same TS segment by tuning the nanowire close to a TS-NTS phase
 transition. \cite{long-PRB}
Bringing the nanowire close to the TS-NTS transition, 
decreases the gap of the system which in turn
 increases the localization length $\xi$  of MFs in the wire and allows
 the tunneling between the initially localized MFs at the ends of TS segments.
The quantitative details of how the tunneling is controlled in topological nanowires 
is discussed in the appendix.

Tunneling of ordinary fermions such as electrons can be used to move electrons from one quantum 
dot to another in a system of quantum dots. In this paper, we will show that the same principle 
applies to MFs, and repeated use of the tunneling Hamiltonian in 
Eq.~\ref{tunneling} can be used to exchange MFs $\gamma_1$ and $\gamma_2$ in a system of TS nanowires 
that hosts such MFs at its ends. 
The unitary time-evolution operator $U$ associated with the exchange  maps 
 $\gamma_1\rightarrow U\gamma_1 U^\dagger=\lambda \gamma_2$
 and  $\gamma_2\rightarrow U\gamma_2 U^\dagger=-\lambda \gamma_1$ where $\lambda$ can be 
directly computed from the tunneling matrix elements $\zeta_{i,j}$ involved in moving the MF $\gamma_1$ to 
the starting position of the MF $\gamma_2$. 
In the low-energy subspace of MFs, the time-evolution operation $U$, that describes the exchange process  
has the usual form of a braid matrix \cite{Ivanov}  
\begin{equation}
U=e^{\frac{\pi}{4} \lambda\gamma_1\gamma_2}.\label{braid}
\end{equation}
While there are only two possible answers $\lambda=\pm 1$ for the braid-matrix, it is critical to be able 
to determine the factor $\lambda$ for a given braid since this is what distinguishes 'clock-wise' 
from 'counter-clock-wise' exchanges.
For the specific geometry discussed in the appendix with wires placed in a superconducting film together with 
 an in-plane magnetic field at $45$ degrees to the wires, the sign of the braid matrix, $\lambda$, 
is determined by the sign of the Rashba spin-orbit coupling constant $\alpha$. 
 
The braiding scheme we will discuss is potentially related to
 measurement-only schemes for braiding of topological quasiparticles. \cite{Parsa-Private,measurementonly} 
However, it is not clear how the anyon model postulates assumed in the measurement-only theory
 apply to the superconducting nanowire systems described by mean-field BCS theory.
For example, the identification of the tunneling matrix element between MFs in Eq.~\ref{tunneling} with the 
topological charge measurement in Ref. \onlinecite{measurementonly} becomes subtle in cases
 where the sign of the tunneling $\zeta_{12}(x)$ oscillates in sign with the separation $x$.
On the other hand the approach in this paper is based only on the  MF tunneling Hamiltonian Eq.~\ref{tunneling}, which 
can be derived microscopically from BCS Hamiltonians. \cite{meng}
The tunneling matrix elements $\zeta_{ij}(x)$
themselves depend on the details of the nanowire system such as the spin-orbit coupling, 
the orientation of the wire and the Zeeman splitting.
Therefore,  we first consider exchange of MFs around a specific triangular loop geometry
in terms of the tunnel matrix elements between the various MFs and then in the appendix, we 
show how the microscopic tunneling parameters may be calculated in one specific geometry.
As a result of our calculation, we find that for general values of the tunneling, the parameter $\lambda$ 
in the braid matrix $U$ has the simple form $\lambda=\textrm{sgn}(\zeta_{12})\chi$ where $\chi$ is the junction 
chirality of the triangular loop that we will define as the product of tunnelings around the loop.
Finally, we would like to note, that while the motivation of exchanging MFs is to be able to 
manipulate the information contained in topological qubits constructed from MFs in an 
effort to perform TQC, it is well-known that braiding by itself is insufficient for TQC.
\cite{bravyi} However, 
MF exchanges are still crucual as one of the most direct tests of non-Abelian statistics and 
 probably  also for any future TQC schemes using MFs. 

\begin{figure}
\subfigure{
\includegraphics[width=.8\columnwidth]{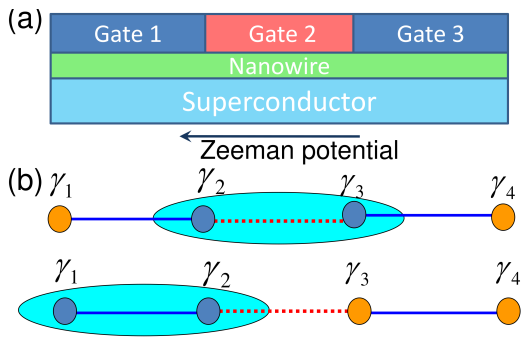}}
\caption{(Color online)(a) Combination of SC, Zeeman and gate potentials
leads to nanowire segments in TS and NTS phases. The gates $1$ and $3$
are adjusted such that the nanowire is in TS phase, while $2$ is adjusted
so that the wire is in NTS phase corresponding to the schematic in
panel (b).
(b) Nanowire segments in TS phase are shown as blue (solid) lines and
 NTS segments are shown as red (dotted) lines. Orange (light) and blue (dark) circles indicate unpaired and paired
 MFs at TS-NTS interfaces, respectively. MFs are paired by tunneling
across the TS or NTS segments denoted by light blue oval.
 Decreasing the tunneling amplitude between
 $\gamma_2$ and $\gamma_3$ and simultaneously increasing
the tunneling amplitude between $\gamma_1$ and $\gamma_2$ can
effectively transfer MF $\gamma_1\rightarrow\gamma_3$.
 }
\label{Fig1a}
\end{figure}

\section{MF transport}To understand how MF transport in a system of
nanowires can be induced by tunneling, consider the simple system of
 nanowires shown in Fig.~\ref{Fig1a}(a) consisting of three semiconducting
 nanowire segments. Two of these segments (shown by solid blue lines in the schematic in Fig.~\ref{Fig1a}(b))
 are in the TS phase, while the wire shown
with the red dashed line is in the NTS phase and serves as the tunnel barrier 
connecting MFs. Each end of the wires in
the TS phase supports a MF (shown as discs). In the initial state
(shown in the upper panel of Fig.~\ref{Fig1a}(b)), the gate voltage of the
NTS segment is chosen to allow a finite tunneling amplitude (shown by light blue oval) across it.
This pairs up the MFs $\gamma_2$ and $\gamma_3$ into finite energy
states with a gap. Thus the operators $\gamma_{2,3}$ become gapped
MFs (shown as dark blue discs) and cannot be used to store quantum
information as can be done with the true zero-energy MFs (shown as light
orange discs).
 The transfer of the MF from position $1$ to $3$ is achieved by
 adiabatically deactivating the tunneling in the NTS segment $2-3$
 and activating the tunneling in the TS segment $1-2$.

 The process of transferring the MF from position $1$ to $3$ shown in Fig.~\ref{Fig1a}(b) 
is described by the time-dependent tunneling Hamiltonian that is derived by extending the 
tunneling Hamiltonian in Eq.~\ref{tunneling} and can be written as  
\begin{equation}
H=[\zeta_{12}\alpha(t)\gamma_1\gamma_2 +\zeta_{23}(1-\alpha(t))\gamma_2\gamma_3]\label{eq:3maj}
\end{equation}
where $\zeta_{12}$ and $\zeta_{23}$ are the activated
tunneling amplitudes across the segments $1-2$ and $2-3$ respectively.
Over the transfer process $\alpha(t)$ varies adiabatically from
 $\alpha(0)=0$ to $\alpha(t_1)=1$.
 It is convenient to understand the
braiding procedure for MF operators
in the Heisenberg representation $\gamma_j(t)=U^\dagger(t)\gamma_j U(t)$
where the $U$ is the unitary time-evolution operator
$U(t)=T e^{-i \int_0^t H(\tau)d\tau}$
 (which is a time-ordered exponential).
  The operators
 $\gamma_j(t)$ can be computed
 from the Heisenberg equation of motion
 $\dot{\gamma}_j(t)=i[H^{(H)}(t),\gamma_j(t)]$.
The Hamiltonian (Eq.~\ref{eq:3maj}) describing the evolution of
 $\gamma_j(t)$ can be written compactly
in terms of an effective B-field $(B_j(t))$ as
\begin{equation}
H^{(H)}(t)= \sum_{a,b,c=1,2,3} \epsilon_{abc} B_a(t) \gamma_b(t)\gamma_c(t)
\end{equation}
where $\epsilon_{abc}$ is the anti-symmetric Levi-Civita tensor.
The time-dependent $B$-field given by
\begin{equation}
\bm B(t)=(1-\alpha(t))\zeta_{23}(1,0,0)+\alpha(t)\zeta_{1,2}(0,0,1).
\end{equation}
The Heisenberg equation of motion for $\gamma_a(t)$ takes the form
\begin{equation}
\dot{\gamma}_a=2 \epsilon_{abc}B_b(t)\gamma_c(t)\label{eq:HOM}.
\end{equation}
This equation of motion
 is identical to that of the spin operators $\sigma_a(t)$ of
a spin-1/2 particle in a  time-dependent magnetic
field $\bm B(t)$
(with a Hamiltonian $H^{(H)}(t)=-\bm B(t)\cdot\bm \sigma(t)$).
 Furthermore, the initial condition on the operator $\gamma(t)=\gamma_1$
corresponds to the spin-operator $\sigma(t)=\sigma_1(0)$ in an initial
effective magnetic field $\bm B(0)=\zeta_{23}(1,0,0)$ that is aligned or
anti-aligned with $\sigma_1$. Thus, after a time-evolution under an
adiabatically varying magnetic field, the spin (and correspondingly the
 MF) remains aligned or anti-aligned with
the final magnetic field  $\bm B(t_1)=\zeta_{12}(0,0,1)$ at time $t=t_1$.
This leads to the expression
\begin{equation}\label{eq:mfmotion}
\gamma_3(t_1)=\textrm{sgn}(\zeta_{12}\zeta_{23})\gamma_1(0).
\end{equation}
Thus the transfer of the tunneling amplitude from the segment $2-3$
 to the segment $1-2$, leads to transport
of the MF from position $1$ to position $3$.
 The hopping of MFs between sites described by Eq.~\ref{eq:mfmotion} is identical to the
motion of regular fermionic operators under that action of tunneling.
We will represent this process by the MF trajectory
 \begin{equation}1\overset{2}{\longrightarrow} 3.\label{eq:arrownotation}\end{equation}
The result in Eq.~\ref{eq:mfmotion} is consistent with a somewhat
 different approach suggested by
 Kitaev. ~\cite{Kitaev-Private}

\section{MFs as defects in dimer lattices}
\begin{figure}
\subfigure{
\includegraphics[width=.8\columnwidth]{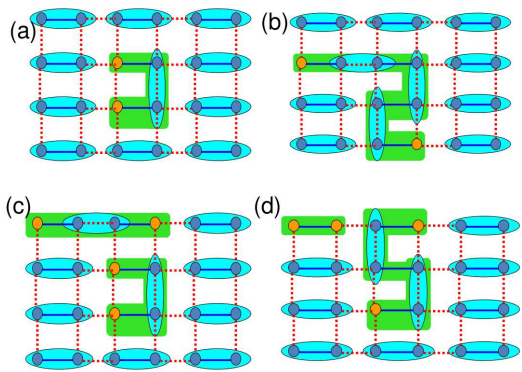}}
\caption{(Color online)(a) Configuration of TS nanowires 
in a square dimer lattice with 2 isolated MFs. MFs are effectively 
bound to defects (unpaired sites) on the dimer lattice. Tunneling between 
different TS segments fuses TS wires into isolated 
effective 'topological wires' shown in green (gray) blocks with MF end modes.
(b) MF transfer processes analogous to Fig.~\ref{Fig1a} can extend effective 
TS wires and move MFs in dimer lattice in a way analogous to Alicea et al.\cite{alicea1}.
(c) and (d): Similar transfer processes can be used to switch the MF end modes between different 
effective 'topological wires'. All operations of the MF square lattice either exchange, contract or 
switch ends of different 'topological wires'.
 }
\label{Fig1b}
\end{figure}
In the previous section, we saw that to transport a single MF from one position to another
 it was necessary to make use of another pair of MFs which were coupled by a weak tunneling so 
that they were not strictly zero-energy MFs. The single MF being exchanged had to be a 
true zero-energy MF with no tunneling, while the pair of MFs with tunneling between them 
may be thought of as a gapped MF dimer. Exchanges of true zero-energy MFs requires a generalization of this 
picture to include several isolated MFs which are not coupled to any other MF by tunneling. 
As is clear from Fig.~\ref{Fig1a}(b), such a process also requires a supply of gapped pairs of MFs (i.e.
MF dimers). In this paper, we will consider a system of nanowires with end MFs most of which are 
paired up by tunneling into MF dimers as shown in Fig.~\ref{Fig1b}(a).
 If all MF sites are completely paired up, then the system has 
no true zero-energy MFs and no topological degeneracy or non-Abelian statistics. Therefore we consider 
a system, where in addition to the MF dimers, there are a few isolated MFs that are unpaired by 
tunneling. If one considers a regular lattice of nanowires so the the end MFs live on the vertices of 
 a square lattice (Fig.~\ref{Fig1b}),
  a system of dimerized MFs forms a dimer covering of the lattice, while isolated 
MFs are associated with defects (unpaired sites) in the MF dimer lattice. As seen by comparing 
Fig.~\ref{Fig1b}(a) and (b), the transport of MFs on the dimer-lattice
is obtained by changing the dimerization pattern on the dimer lattice analogous to Fig.~\ref{Fig1a}(b).

The MFs on the dimer lattice can also be thought of as the end points of topological wires. To see 
this we define effective 'topological wires', shown by green (gray) boxes in Fig.~\ref{Fig1b}, by 
considering TS wire segments coupled by tunneling as a single topological wire. This identification, 
relates our proposal in a direct way to the proposal of Alicea et al.\cite{alicea1} However, 
the details of the physical implementation remain different and the dimer implementation
presented in this paper will allow us to directly use Eq.~\ref{eq:mfmotion}, to determine the 
form of the braid-matrix in Eq.~\ref{braid}. The continuous processes required by 
Alicea et. al. for exchanging MFs were extending and contracting topological wire segments  together 
with an operation that we will refer to as exchanging the ends of different topological wires. 
This process required bringing together a pair of topological wires in a tri-junction and effectively 
takes a pair of topological wires with end points $\gamma_{1,2}$ and $\gamma_{3,4}$ and 
creates a new pair of wires with end points $\gamma_{1,3}$ and $\gamma_{2,4}$. All these processes 
can be accomplished in MF dimer lattice by repeated application of the process shown in Fig.~\ref{Fig1a} 
associated with Eq.~\ref{eq:mfmotion}. The analogue of the extension and contraction process in a dimer 
lattice is shown in Fig.~\ref{Fig1b}(a) and (b), while the end switching process is shown in   Fig.~\ref{Fig1b}(c) and (d).

\section{Non-Abelian statistics of MFs in nanowires}
In this section, we show explicitly, that exchange of MFs in any dimer lattice can always be 
described by an equation of the form of Eq.~\ref{braid}.
Unpaired MFs can be exchanged via  discrete  tunneling operations of the form shown in Fig.~\ref{Fig1a}(b).
Since the physical positions of the MFs are exchanged by the correct sequence of MF transfers, 
 the resulting transformation of the MFs
at the end of the transformation $t=t_{final}$ has the general form
\begin{align}
\gamma_1(t_{final})= \lambda\gamma_2(0)\nonumber\\
\gamma_2(t_{final})= \tilde{\lambda}\gamma_1(0).\label{eqlambda}
\end{align}
However, consistency with non-Abelian statistics also require us to
prove that $\lambda\tilde{\lambda}=-1$. If $\lambda\tilde{\lambda}=-1$,
the exchange transformation can be represented by the operator $U$ of the form Eq.~\ref{braid}.

To show $\lambda\tilde{\lambda}=-1$, let us label the unpaired MFs,
which are to be exchanged as  $1$ and $2$ and take all other MFs
as paired, $(2 n-1,2 n)$  for $n=2,\dots,N$. The positions of the MFs following each step (labelled by the index $p$)
of the exchange process, which permutes the positions of the MFs, can be
 represented by the function $\pi_p(j)$, where $j=1,\dots,2N$ is the MF index.
After each step $p$, the MF coordinates are updated from $\pi_{p-1}$  to $\pi_p$ according to the relation
\begin{equation}
\pi_p(j)=\pi_{p-1}(C_p(j))
\end{equation}
where $C_p$ is a cyclic (clockwise or anticlockwise) permutation of the MFs $a_p,\,2 n_p-1$ and
 $2 n_p$ corresponding to Eq.~\ref{eq:mfmotion}.
Here we choose $a_p$ to be one of the unpaired MFs, $1\textrm{ or }2$,
and $n_p>1$ such that  MF dimer $(2 n_p-1,2 n_p)$ is paired.
The equation of motion for the unpaired Majorana operators corresponding to Eq.~\ref{eq:mfmotion} is
\begin{equation}
\gamma_{\pi_{p+1}(a_p)}(t_{p+1})=\lambda_p\gamma_{\pi_p(a_p)}(t_p).
\end{equation}
where
\begin{equation}
\lambda_p=\textrm{sgn}(\zeta_{\pi_{p+1}(2 n_p-1)\pi_{p+1}(2 n_p)}\zeta_{\pi_{p}(2 n_p-1)\pi_{p}(2 n_p)}).\label{eqlambdap}
\end{equation}
The total sign $\lambda\tilde{\lambda}$ picked up by
the unpaired MFs is the product
 $\lambda\tilde{\lambda}=\prod_p\lambda_p$.
To calculate this product we define a sequence
 $Q_p=\textrm{sgn}(\prod_{n>1}\zeta_{\pi_p(2 n-1),\pi_p(2 n)})$.
 From Eq.~\ref{eqlambdap} it follows
 that $Q_{p+1}=\lambda_p Q_p$
so that
\begin{equation}
Q_{final}=\lambda\tilde{\lambda}Q_{p=0}.
\end{equation}
Note that since each cyclic permutation $C_p$ contains even number of exchanges (i.e is an even permutation), the
permutations $\pi_p$ at each step (including the final permutation
$\pi_{final}$), which is a product of $C_p$s, is also an even permutation.

Since the Hamiltonian is required to return to its initial configuration, the MFs at positions $(2 n-1,2 n)$ must be paired
by tunneling for $n>1$. This requires that $\pi_{final}$ is  composed of a pair exchange of the positions of MF dimers
 $(2 n-1,2 n)\leftrightarrow (2 n'-1,2 n') $ together with possible internal flips $(2n-1\leftrightarrow 2n)$ of the dimers.
 Since, $\pi_{final}$ is an even permutation, and dimer exchanges are even permutations,  the number of
internally flipped dimers  $(2n-1\leftrightarrow 2n)$ in $\pi_{final}$ is even.
 Moreover, the unpaired pair of MFs $(1,2)$ is flipped in $\pi_{final}$.  Thus an odd number
of the paired MF dimers $(2n-1\leftrightarrow 2n)$, must be flipped
for $n>1$. Each such dimer flip changes the sign of $Q_{final}$, since $\zeta_{2 n-1, 2n}=-\zeta_{2 n, 2n-1}$ for $n>1$.
This leads to the relation $Q_{final}=-Q_{p=0}=\lambda\tilde{\lambda}Q_{p=0}$, proving the consistency condition for non-Abelian
statistics i.e. $\lambda\tilde{\lambda}=-1$.

\section{Exchange around a triangular loop}
\begin{figure}
\includegraphics[width=.8\columnwidth]{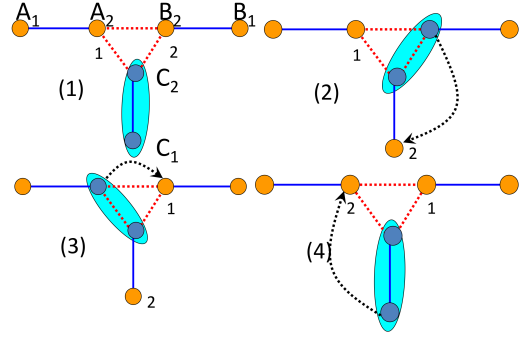}
\caption{(Color online)
 MFs $1$ and $2$  at the ends of different
TS segments are exchanged. This is achieved  by switching
tunnelings on and off on TS and NTS segments in 4 steps going from
a state shown in one panel to the next panel. Dotted arrow shows motion
of MF from the previous panel.The labelling for the
sites $A_1$,$A_2$, $B_1$,$B_2$, $C_1$ and $C_2$ is shown in panel (1).}
\label{Fig3}
\end{figure}

\begin{figure}
\includegraphics[width=.8\columnwidth,angle=270]{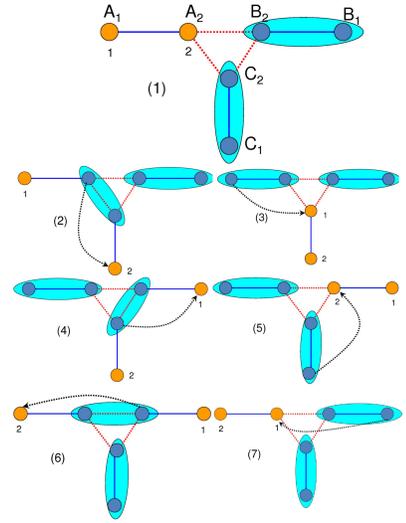}
\caption{(Color online) MFs $1$ and $2$
 at the ends of the TS segment on the left leg are exchanged in seven steps
similar to Fig.~\ref{Fig3}.
 Step (7) transfers state shown in panel (6) back to panel
 (1) with the effect that the Majoranas 1 and 2 are interchanged. }
\label{Fig2}
\end{figure}

A specific realization of the non-Abelian statistics in nanowire
systems is provided by a triangular loop
geometry shown in Figs.~\ref{Fig3} and ~\ref{Fig2}.
The triangular loop consists of one end ($A_2$,$B_2$ and $C_2$) of each of
 three TS segments (A, B and C) connected by NTS segments to form
a triangle. The other ends are labeled  $A_1$, $B_1$ and $C_1$.
 The MFs to be exchanged,
 referred as $1$ and $2$, are assumed to be  localized at
2 of these 6  ends of TS segments.
Each of the steps for the MF exchange (shown in Figs.~\ref{Fig3} and
~\ref{Fig2}  )
 consists of moving exactly one MF from one position to
 the other (shown by dotted arrows) by adiabatically turning off the
 tunneling in some wire segment and
increasing it in an adjoining segment as discussed before.

The procedure to exchange the MFs $1$ and $2$ at the ends of different TS
 segment through the tri-junction takes place
 in four steps shown in Fig.~\ref{Fig3}.
The signs associated with the exchange $\lambda$ and $\tilde{\lambda}$
can be determined by following the trajectories of the MFs
$1$ and $2$ and applying Eqs.~\ref{eq:mfmotion} and ~\ref{eqlambda}.
From Fig.~\ref{Fig3}, it is clear that the sequence of positions
 followed by the MFs $1$ and $2$ are
\begin{align}
&\textrm{MF 1:  }A_2\overset{C_2}{\underset{(3)}{\longrightarrow}} B_2\nonumber\\
&\textrm{MF 2:  }B_2\overset{C_2}{\underset{(2)}{\longrightarrow}} C_1\overset {C_2}{\underset{(4)\equiv (1)}{\longrightarrow}} A_2\label{eqdiff}
\end{align}
respectively.
Here we show only the MF that is moved in each step, which is numbered
in Fig.~\ref{Fig2} as $(j=1,\dots,4)$ (marked below the arrows in Eq.\ref{eqdiff}). The MF motion is
shown using the notation defined in Eq.~\ref{eq:arrownotation} so that the sign can be calculated using Eq.~\ref{eq:mfmotion}.
Applying Eq.~\ref{eq:mfmotion}, the parameters $\lambda$ and
 $\tilde{\lambda}$ simplify to
\begin{equation}
\lambda=-\tilde{\lambda}=\textrm{sgn}(\zeta_{A_2 B_2})\chi\label{eqlambda2}
\end{equation}
where
 $\chi=\textrm{sgn}(\zeta_{A_2 B_2}\zeta_{B_2 C_2}\zeta_{C_2 A_2})$ is defined to be the
chirality of the tri-junction.\cite{david}

Similarly MFs at the ends of the same TS segment
can be exchanged using six steps shown in Fig.~\ref{Fig2}.
From Fig.~\ref{Fig2}, it is clear that the sequence of positions followed
by the MFs $1$ and $2$ are
\begin{align}
&\textrm{MF 1:  }A_1\overset {A_2}{\underset{(3)}{\longrightarrow}} C_2\overset {B_2}{\underset{(4)}{\longrightarrow}} B_1\overset {B_2}{\underset{(7)\equiv (1)}{\longrightarrow}} A_2\nonumber\\
&\textrm{MF 2:  }A_2\overset {C_2}{\underset{(2)}{\longrightarrow}} C_1\overset {C_2}{\underset{(5)}{\longrightarrow}} B_2\overset {A_2}{\underset{(6)}{\longrightarrow}} A_1
\end{align}
respectively. The step $(7)$ is not explicitly shown in Fig.~\ref{Fig2}, since it is equivalent to $(1)$.
Applying Eq.~\ref{eqlambda}, the parameters
 $\lambda$ and $\tilde{\lambda}$ simplify to
\begin{equation}
\lambda=-\tilde{\lambda}=\textrm{sgn}(\zeta_{A_1 A_2})\chi\label{eqlambda1}
\end{equation}
where $\chi$ is the junction chirality.

Thus, using Eq.~\ref{eqlambda1} and Eq.~\ref{eqlambda}, we obtain the
 the result that  the unitary time-evolution of the MFs
  $\gamma_1$ and $\gamma_2$ under exchange can be described by the unique braid-matrix
\begin{equation}\label{eqbraid2}
U=e^{\frac{\pi}{4}\chi \textrm{sgn}(\zeta_{12})\gamma_1\gamma_2}
\end{equation}
where $\zeta_{12}$ is the tunneling amplitude of the segment
separating $\gamma_1$ and $\gamma_2$.
 The quantities $\zeta_{12}$ and $\chi$
for a specific network are calculated
 in the appendix.

\section{Conclusion}
Non-Abelian statistics for MFs at the ends of
 TS nanowire segments can be realized by introducing
time-varying gate-controllable
tunnelings between  MFs in a nanowire system to exchange the end MFs.
Similar to the previous proposal for braiding MFs in 1D wires, our system can 
also be embedded in 3D leading to the possibility of non-Abelian statistics in 3D.
 The isolated MFs being exchanged in the 
tunneling geometry considered in this paper may be thought of as defects in a dimer 
lattice i.e. sites that are unpaired by tunneling. Alternatively, this system may also be 
thought of as a discretized implementation of the continuous nanowire network proposal of 
Alicea et. al.\cite{alicea1} However, the discrete implementation discussed in this paper 
allows us to compute the braid matrix explicitly in terms of MF overlaps. 
 The non-Abelian braid matrix
for exchange around a triangular loop geometry is given by a product 
of the fusion channel of the MFs $\zeta_{1,2}$ and the junction-chirality $\chi$.
The fusion channel $\zeta_{1,2}$ is simply the tunneling matrix-element between the MFs being 
exchanged and the junction-chirality is the product of tunneling 
terms around the triangular junctions. Thus the braid-matrix in the tunneling geometry 
considered in this paper is completely determined in terms of microscopic
tunneling parameters  by Eq.~\ref{eqbraid2} making nanowire systems
 a well-controlled platform to realize non-Abelian statistics. 

\acknowledgements{We thank Parsa Bonderson, Anton Akhmerov and Kirill Shtengel for helpful discussions. We are grateful to the Aspen Center for Physics for hospitality during the 2010 summer program \emph{Low Dimensional Topological Systems}.  D.J.C. is supported in part by the DARPA-QuEST program. S.T. acknowledges support from DARPA-MTO Grant No: FA 9550-10-1-0497. J.D.S. is supported by DARPA-QuEST, JQI-NSF-PFC, and LPS-NSA.}

\appendix
\section{Calculation of tunneling matrix elements for a specific 
nanowire system}
\maketitle
In the main text of the paper we saw how controlling the tunneling 
between MFs can be used to generate transport of Majorana fermions 
from one point to another and eventually generate exchanges and 
braids that are useful for TQC. The sign of the resulting 
exchange was found to be determined by the signs of various tunneling 
matrix elements. While the existence of non-Abelian statistics is 
demonstrated in the paper in general, the signs of the tunneling 
matrix elements themselves are depend on the microscopic details 
of the system. 

\begin{figure}
\subfigure{
\includegraphics[width=.8\columnwidth]{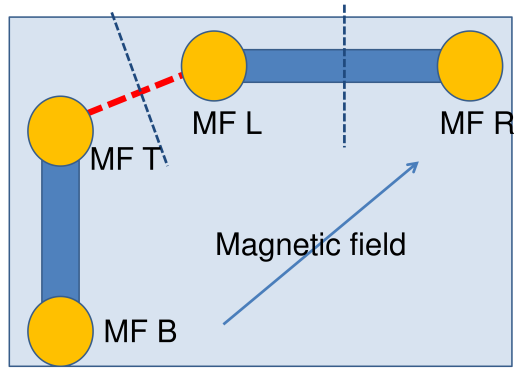}}
\caption{(Color online)Schematic of orthogonal nanowire system on a 
superconductor (shown as light rectangle) that generates tunneling
 of MFs (shown as light orange discs). The entire system is subject to 
an in-plane magnetic field to generate Zeeman coupling. 
 Nanowire segments in the TS phase are shown as dark blue rectangles with 
end MFs. Tunneling is generated between MF T and MF L by conventional 
tunneling across a nearly depleted nanowire in the NTS phase. The tunneling can be
 calculated using the Bardeen tunneling formula \cite{bardeen} as 
the matrix element of the current operator in the middle of the wire (black dotted line).
 Similarly tunneling is generated between MF L and MF R by lowering the topological gap
 so that the wave-functions have significant overlap at the middle of the wire (black dotted line). 
 }
\label{Figepaps}
\end{figure}

In this appendix, we calculate the tunneling matrix elements 
 between the various MFs at a junction for a network of orthogonal
 wires on a superconducting
 substrate as shown in Fig.~\ref{Figepaps}.
A Zeeman potential is applied at 45 degrees to the wires. The wires
 in the TS phase (shown in dark blue), which support MFs (shown as orange 
discs) at their ends,
 are taken to have a Rashba spin-orbit coupling generated from interaction 
with the superconducting substrate.
The Bogoliubov de Gennes (BdG) Hamiltonian for the wire along $x$ is
 given by
\begin{equation}
H_{BdG}=(-\eta \partial_x^2-\mu(x))\tau_z + V_z\bm\sigma\cdot \hat{\bm B}+\imath\alpha\partial_x \sigma_y\tau_z+\Delta\tau_x\label{eq:BdG}
\end{equation}
and for the wire along $y$ is given by
\begin{equation}
H_{BdG}=(-\eta \partial_y^2-\mu(y))\tau_z + V_z\bm\sigma\cdot \hat{\bm B}-\imath\alpha\partial_y \sigma_x\tau_z+\Delta\tau_x.
\end{equation}
The direction of the Zeeman field is $\hat B=(\hat{x}+\hat{y})/\sqrt{2}$.
 Following the
spin-rotation and phase transformations in Ref.~[\onlinecite{long-PRB}],
 for negative Rashba coupling $\alpha<0$, the Majorana wave-functions at the left and the right ends of the
 $x$ wires have the form
\begin{align}
&\phi_L=\left(\begin{array}{c}u(x)e^{\imath \phi/2}\\ \imath\sigma_y u(x)e^{-\imath \phi/2}\end{array}\right) \textrm{ and } \phi_R=\imath\sigma_x\left(\begin{array}{c}u(-x)e^{-\imath \phi/2}\\ \imath\sigma_y u(-x)e^{\imath \phi/2}\end{array}\right)
\end{align}
respectively,where $\phi=\sin^{-1}{\frac{V_Z}{\Delta\sqrt{2}}}$ and $u(x)$
is a real 2-spinor. Note that in this geometry there is now an additional condition for the wire to be gapped i.e
$\sqrt{\Delta^2+\mu^2}<V_Z<\Delta\sqrt{2}$.
This constraint implies $\frac{\pi}{4}<\phi<\frac{\pi}{2}$.
 The Majorana wave-functions for the Majorana fermions
at the bottom and top ends of the wires parallel to the $y$ axis
  have the form
\begin{align}
&\phi_B=Q\left(\begin{array}{c}u(y)e^{-\imath \phi/2}\\ \imath\sigma_y u(y)e^{\imath \phi/2}\end{array}\right) \textrm{ and } \phi_T=\imath Q\sigma_x\left(\begin{array}{c}u(y)e^{\imath \phi/2}\\ \imath\sigma_y u(y)e^{-\imath \phi/2}\end{array}\right)
\end{align} respectively, where $Q=e^{-\imath \pi\sigma_z/4}$.

Transport of MFs is generated by introducing tunneling into the 
system of MFs shown in Fig.~\ref{Figepaps}. The 
junction chirality $\chi$ defined in the paper depends only  on the tunneling from 
3 end MFs MF $\{L,R,T\}$ with wave-functions $\phi_{L,R,T}$. Let us start by considering the MF overlap across the 
NTS segments (shown as red dashed lines in Fig.~\ref{Figepaps}) 
which is simplest to understand in the limit of low negative chemical 
potential $\mu=-|\mu|$ where $|\mu|\gg V_Z,\Delta$. Physically, this 
corresponds to a wire that is nearly depleted of electrons and only 
acts as a tunnel barrier. In such a case, the Majorana wave-function in the
 NTS wire has the usual exponentially decaying form
 $\Psi(x)=\Psi(x_I)e^{-\gamma |x-x_I|}$ as in a barrier, where $\gamma\sim\sqrt{2 m |\mu|}$
 and $x_I$ is the position of the interface between  the TS and NTS wire 
segments. 
The tunneling matrix elements $\zeta_{ij}$ between two MFs at 
$x_{I,1}=-a/2$ and $x_{I,2}=a/2$ across the NTS wire can be
calculated from the matrix-elements of the current operator and the 
wave-functions in the middle of the wire \cite{bardeen} ($x=0$ as shown by the dark dotted lines in Fig.~\ref{Figepaps})
\begin{align}
&\zeta=\frac{1}{2 }[\Psi^\dagger_1(0)\tau_z \partial_x\Psi_2(x)|_{x=0}-\partial_x\Psi^\dagger_1(x)|_{x=0}\tau_z\Psi_2(0)]\nonumber\\
&-i\alpha \Psi^\dagger_1(0)\sigma_y\tau_z\Psi_2(0)\sim - \gamma e^{-\gamma a}\Psi^\dagger_2(-\frac{a}{2})\tau_z\Psi_1(\frac{a}{2})\nonumber\\
&=- \rho \Psi^\dagger_2(-\frac{a}{2})\tau_z\Psi_1(\frac{a}{2})
\end{align}
where $\rho=\gamma e^{-\gamma a}$ is the overall tunneling strength and we have assumed $\lambda\gg \alpha$.
  In this limit, the overlap between a pair of Majorana wave-functions
 $\Psi_1=(u_1(x),\imath\sigma_y u_1^*(x))^T$ and $\Psi_2=(u_2(x),\imath\sigma_y u_2^*(x))^T$
is given by $M= 2\imath \rho Im(u_1^\dagger u_2)$. This is purely imaginary and
manifestly anti-symmetric as expected. Furthermore since the fundamental
spinor $u(x)$ in terms of which each of $u_{1,2}$ are written is real,
we can write it as $u=(\cos{\theta},\sin{\theta})^T$, where the parameter $\theta$
 depends on $V_Z,\mu,\alpha$ etc.
 With the help of these relations
it is easy to tabulate the Majorana tunneling matrix as an anti-symmetric
matrix for the states in the order $(L,R,T)$ as
\begin{equation}
\zeta=\imath\frac{\rho}{\sqrt{2}} \left(\begin{array}{ccc}0& \sqrt{2}\cos{\phi}\sin{2\theta} &  \sin{2\theta}\\
*&0& (\sin{\phi}+\cos{\phi}\cos{2\theta}) \\
*&*&0\end{array}\right)\label{Smatrix}
\end{equation}
where the elements in the $*$ have been left empty since they are
 determined by the anti-symmetry constraint.
The junction chirality $\chi$  in the previous section, used only the Majorana modes $L,T,R$ and is calculated
 using the expression,
$\chi=\zeta_{RL}\zeta_{LT}\zeta_{TR}=\rho^3\cos^2{\phi}\sin^2{2\theta}[\cos{2\theta}+\tan{\phi}]$, which is always positive,
 since $\tan{\phi}>1$ for the Zeeman direction   $\hat B=(\hat{x}+\hat{y})/\sqrt{2}$ and  negative Rashba coupling $\alpha<0$.
Changing the Zeeman potential to  $\hat B=(\hat{x}-\hat{y})/\sqrt{2}$
 flips the chirality. Changing the sign of the Rashba coupling $\alpha$
requires us to change $\phi_L\rightarrow\imath (u(x)e^{\imath\phi/2},-\imath\sigma_y u(x)e^{-\imath\phi/2})$. Since all the other wave-functions are
derived from symmetry transformations applied to $\phi_L$, the rest
of the calculation goes through as is with the only difference that $u(x)$
changes to $\imath u(x)$. Therefore the final result for the chirality
of the junction is independent of the Rashba coupling.

The signs acquired by MFs on exchange is dependent also on the tunneling between the MFs MF L,R across a TS segment.
 The tunneling amplitude between Majorana
 fermions on the same topological
segment is well-controlled and can be calculated in the limit of a long topological wire
 (wire length $L> \alpha/V_Z$). In this limit, the MFs only overlap
in the limit where the gate potential is tuned so that the wire is driven towards a phase
 transition by tuning $\mu$ near
 to $\mu=\sqrt{V_Z^2-\Delta^2}$.
The relevant slowest decaying spinor component then determines the
 tunneling matrix elements and is given by  $u(x)=(V_Z+sgn(\alpha)\sqrt{V_Z^2-\mu^2},-\mu)^T exp(-\frac{x}{|\alpha|} (\sqrt{V_Z^2-\mu^2}-\Delta))$.
 The tunneling matrix element
 is given by $M\sim-i\alpha \Psi_L^\dagger \sigma_y \tau_z\Psi_R=-i\alpha Re[u_L^\dagger \sigma_y u_R]$. Substituting $u$, we find the
overlap to simplify to $M\propto -i\alpha \cos{\phi}e^{-2 L/\xi}$ where $\xi=|\alpha|/(\sqrt{V_Z^2-\Delta^2}-\mu)$.
Thus the sign of $\zeta_{L,R}$ is determined by the sign of the Rashba
 spin-orbit coupling $\alpha$.
Using these results together with Eq.~\ref{Smatrix}, one can check the physically reasonable 
result that changing the sign of the Rashba coupling $\alpha$,
 flips 
 the sign of the MFs on interchange.

Thus, provided care is taken to ensure the conditions for the
braid discussed in this section, the sign of both clock-wise exchanges
 is positive for
positive Rashba coupling $\alpha$ and negative other-wise. For a given
Rasbha coupling the sign of the braid can be altered by considering anti-clockwise braids.

\end{document}